\documentclass[twocolumn,pra,aps,showpacs]{revtex4}
\usepackage{graphicx}
\usepackage{amsmath}
\usepackage{amssymb}

\begin{document}

\title{On the properties of information gathering in quantum and classical measurements}

%\title{Structural properties of information gathering in Q and C measurement}
%\title{On the structure of information gathering in quantum and classical measurement}
%\title{On the information theoretic structure of quantum and classical measurement}

%\title{On the strength and classical capacity of quantum measurements}
%\title{Some properties of measurements: strength, disturbance and classical capacity}
%\title{Measurement strength, disturbance and classical capacity}

\author{Kurt Jacobs}

\affiliation{Centre for Quantum Computer Technology, Centre for 
Quantum Dynamics, School of Science, Griffith University, Nathan 4111, 
Brisbane, Australia}

\begin{abstract}
The information provided by a classical measurement is unambiguously 
determined by the mutual information between the output results and 
the measured quantity.  However, quantum mechanically there are at 
least two notions of information gathering which can be considered, 
one characterizing the information provided about the initial 
preparation, useful in communication, and the other characterizing the 
information about the final state, useful in state-preparation and 
control.  Here we are interested in understanding the properties of 
these measures, and the information gathering capacities of quantum 
and classical measurements.  We provide a partial answer to the 
question `in what sense does information gain increase with initial 
uncertainty?'  by showing that, for classical and quantum measurements 
which are symmetric with respect to reversible transformations of the 
state space, the information gain regarding the initial state always 
increases with the observer's initial uncertainty.  In addition, we 
calculate the capacity of all unitarily covariant and commutative 
permutation-symmetric measurements for obtaining classical 
information.  While it is the von Neumann entropy of the effects which 
appears in the latter capacity, it is the subentropy which appears in 
the expression for the former.
\end{abstract}


\maketitle

\section{Introduction}
In this article we will be concerned with, among other things, the 
question `what is the capacity of a given measurement to provide 
information?'  In considering this question, it is important to note 
that for quantum measurements the answer is not unique, and depends on 
what the information is about.  Thus, we will also be concerned with 
the question of how information gain can be quantified, and in doing 
so we will present two kinds of information gain useful for different 
applications.  Finally, for a given measure of the information gain, 
we will be interested in asking what properties it has, in particular 
how it depends upon the observer's initial uncertainty.

Since we will be concerned with both classical and quantum 
measurements, we begin in Section~\ref{CandQM} by discussing the 
relationship between the two, and in particular how classical 
measurements may be described as a subset of quantum measurements, 
providing us with a unified description of both.  In 
Section~\ref{MOps} we describe two physically motivated operations 
which can be used to combine measurements, and introduce some 
notation.  In Section~\ref{CCandMS} we discuss two kinds of 
information gathering, information about the initial ensemble, and 
information about the final state, and consider some questions 
regarding their respective properties under operations on 
measurements.  In Section~\ref{Dist} we pause to consider quantifying 
disturbance, and discuss how one can define two measures of 
disturbance corresponding loosely to the two notions of information 
gain considered in the Section~\ref{CCandMS}.  In Section~\ref{IGSSS} 
we consider the information gain regarding the initial preparation, 
and show that there is a precise sense in which it can be said to 
increase with the observer's initial uncertainty.  This motivates the 
definition of two classes of measurements, the permutation-symmetric 
measurements and the unitarily covariant measurements.  Finally, we 
calculate the capacity of these classes to obtain information about 
the initial preparation.

\section{Classical and Quantum Measurements}
\label{CandQM}
The purpose of this section is to describe the relationship between 
classical and quantum measurements, and discuss the special role of 
{\em bare} quantum measurements, the class with which we will be 
primarily concerned throughout.  Quantum 
measurements~\cite{Schumacher,Kraus} are described by sets of 
operators, $\{\Omega_n\}$, which satisfy $\sum_n \Omega_n^\dagger 
\Omega_n = 1$.  For efficient measurements each operator in the set 
corresponds to a possible outcome of the measurement~\cite{introMeas}.  
(Inefficient measurements, on the other hand, are merely efficient 
measurements in which the result is not known with certainty - 
however, we will only be concerned with efficient measurements in what 
follows).  Due to the polar decomposition theorem~\cite{Schatten}, we 
can write each of the operators $\Omega_n$ as a product of a unitary 
and a positive operator.  As a result, the state following a 
measurement may be written as
\begin{equation}
    \rho_n = \frac{U_n Q_n \rho Q_n 
                   U_n^\dagger}{\mbox{Tr}[Q_n^2\rho]} ,
  \label{polarform}
\end{equation}
where $U_n$ is a unitary, and $Q_n$ is positive.  So long as the 
observer has the necessary Hamiltonian resources, she can always 
perform any unitary operation after the measurement, and conditional 
upon the result.  Such an action is often referred to as {\em 
feedback}~\cite{WM,DJ,Belavkin}.  Because of this, we can conclude two 
things from the above expression for the final state resulting from 
the outcome $n$.  The first is that, using unitary feedback, one can 
simulate any quantum measurement by making a measurement whose 
operators $\Omega_n$ are purely positive.  The second is the converse; 
that, given any measurement, one can always use unitary feedback to 
simulate a measurement in which the operators are purely positive.  
However, there is a deeper significance to the polar form given in 
Eq.(\ref{polarform}), but to see this we must turn to classical 
measurements.

 % ********* was at	the	state of `Classical	Measurements'***********
 % As we discussed in Section~\ref{CandQM},	classical measurement theory, 
 % which existed, naturally, long before quantum theory, describes 
 % measurements	performed on classical objects and derives directly	from 
 % Bayesian	statistical	inference.	Quantum	measurement	theory reduces to 
 % classical measurement theory	when only a	single basis is	relevant, 
 % both	to the state of	the	system being measured, and to the measurement 
 % itself.	That is, quantum measurement theory	reduces	to the theory of 
 % Bayesian	statistical	inference when both	the	density	matrix and the 
 % measurement operators are all diagonal in the same basis.  In that 
 % case, the POVM describes	merely a classical measurement being 
 % performed on	a classical	random variable	which has an initial 
 % probability distribution	given by the diagonal of $\rho$.  Thus,	we 
 % need	not	introduce any new formalism	for	classical measurements,	we 
 % need	merely to define them as a subset of quantum measurements.
 % **************************************************************

Classical measurement theory, which existed, naturally, long before 
quantum theory, describes measurements performed on classical objects 
and derives directly from Bayesian statistical 
inference.~\cite{Bayes,Jaynes,ABayesBook}.  While classical 
measurements are not usually written using the same formalism as 
quantum measurements, they can be, and this is to be expected because 
quantum measurements must reduce to classical measurements under 
certain conditions; quantum measurement theory reduces to the theory 
of Bayesian statistical inference when both the density matrix and the 
measurement operators are all diagonal in the same basis.

To describe classical measurements with the same formalism as quantum 
measurements, one writes the observer's initial classical state of 
knowledge (being a probability density) about a classical variable 
$x$, as a positive diagonal matrix, $\rho_{\mbox{\scriptsize c}}$.  
Once one has done this, a classical measurement may be described by a 
set of positive diagonal operators $P_n$, such that $\sum_n P_n^2 = 
1$.  The state that results from the $n^{\mbox{\scriptsize\it th}}$ 
outcome is
\begin{equation}
    \rho_n = \frac{ P_n \rho P_n }{\mbox{Tr}[P_n^2\rho]} .
  \label{classmeas}
\end{equation}
For an explanation of how this form is derived from the theory of 
Bayesian inference, the reader is referred to Ref.~\cite{pool}.  Thus, 
classical measurements are described by a subset of quantum POVM's, in 
which all the operators must be positive, mutually commuting and 
commuting with the `density operator' describing the initial classical 
state of knowledge of the observer.

However, it is useful to note that we can add to the classical 
formalism deterministic transformations.  In the classical case, the 
only deterministic operations which can be performed are permutations 
of the states of the random variable.  In our formalism, these are 
implemented by multiplying the classical (diagonal) density matrix on 
the left and right by permutation matrices.  (Permutation matrices are 
such that every element is zero or unity, and there is only one 
non-zero element in each row and column.)  If we allow the observer to 
perform a deterministic transformation after the measurement, and 
conditional upon the outcome, then we have
\begin{equation}
    \rho_n = \frac{ T_n P_n \rho P_n T_n^\dagger}{\mbox{Tr}[P_n^2\rho]} .
  \label{classOp}
\end{equation}
When we augment classical measurement with classical deterministic 
transformations, one could refer to the result as a {\em classical 
operation}.  While Eq.(\ref{classOp}) is sufficient for our purposes, 
it does not give the most general form for a classical operation.  To 
do this one would include, in addition to the above, a stochastic map 
describing an irreversible randomization of the system.  We note that 
a discussion of the relationship between quantum and classical 
measurements, using a different formalism, is given in~\cite{Hardy}.

There is now a close similarity between the above expression for 
classical operations, and the expression for quantum operations given 
in Eq.(\ref{polarform}).  In fact, this similarity is not just 
superficial; there is a sense in which we can think of the quantum 
positive operators $Q_n$ as characterizing the unadorned act of 
measurement, as the $P_n$ do in classical operations.  To see this we 
consider the von Neumann entropy of the average state of the system 
after the measurement.  Classically, if we allow only measurement, we 
have
\begin{equation}
    \sum_n p_n \rho_n = \sum_n P_n \rho P_n = \sum_n P_n^2 \rho = 
    \rho ,
  \label{classMent}
\end{equation}
so that the entropy of the final average state is the same as the 
entropy of the initial state.  Denoting the difference between the 
initial and final entropy as
\begin{equation}
    D_{\mbox{\scriptsize o}}(\rho) \equiv S\left[ \sum_n p_n \rho_n \right] - S[\rho],
  \label{Do}
\end{equation}
(a notation which will be motivated later) then for classical 
measurements we have $D_{\mbox{\scriptsize o}} = 0$.  However, if we 
allow conditional deterministic transformations (feedback), then it 
becomes possible to arrange that $D_{\mbox{\scriptsize o}} < 0$.  That 
is, feedback allows us to reduce the final average entropy below the 
initial entropy, which is the essence of feedback control.

A similar result exists for quantum measurements, which is due to a 
theorem of Ando~\cite{AndoMaj}.  Ando's theorem states that for 
quantum measurements in which all the operators are positive 
($\Omega_n = Q_n, \;\forall n$), $D_{\mbox{\scriptsize o}}(\rho) \geq 
0$.  In addition, when one allows the use of deterministic feedback 
(unitary operators), as in the classical case one can arrange that 
$D_{\mbox{\scriptsize o}}(\rho) < 0$.  This therefore provides an 
initial quantitative motivation for thinking of the positive operators 
which appear in the polar decomposition for quantum operations as the 
equivalent of the measurement operators in classical operations.  
Moreover, we expect that other relations exist in which quantum 
operations with positive operators parallel classical measurements, 
and we give a conjecture regarding a second example of such a relation 
in Section~\ref{CCandMS}.

In view of the above argument we will refer to quantum measurements in 
which all the measurement operators are positive as {\em bare} 
measurements.  An important feature of bare measurements is that, due 
to the Polar decomposition theorem, if one is interested purely in the 
process of acquiring information, it is enough to study these alone, 
since post-measurement unitary transformations do not affect this 
process.  We note that such measurements have been referred to 
previously by various authors as, {\em pure} measurements~\cite{DJJ}, 
measurements {\em without feedback}~\cite{FJ}, and {\em square root} 
measurements~\cite{Barnum}.  

\section{Operations on measurements}
\label{MOps} 
It is useful at this point to introduce a few definitions.  In what 
follows we will denote measurements by calligraphic letters.  If we 
write,
\begin{equation}
 {\mathcal M} \equiv \{ \Omega_n : n = 1,\ldots, M \}
\end{equation}
we will mean that ${\mathcal M}$ denotes the measurement described by 
the operators $\{ \Omega_n \}$.  The operators $\{ \Omega_n \}$ are 
often referred to as the Kraus operators~\cite{Kraus}, although in 
what follows we will refer to them simply as the measurement 
operators.  We will follow Kraus's terminology and refer to the 
operators $E_n = \Omega_n^\dagger \Omega_n$ as the {\em effects}.

There are at least two ways in which a physical procedure may be used 
to combine two measurements together to form a new measurement.  One 
we will refer to as {\em mixing}~\cite{mixing}, and the other as {\em 
concatenation}.

{\em Mixing:} In this case the observer chooses between two
measurements, ${\mathcal M} \equiv \{ \Omega_m^{\mathcal M} : m =
1,\ldots,M \}$ and ${\mathcal N} = \{ \Omega_n^{\mathcal N} : n =
1,\ldots,N \}$ on a random basis, choosing to use ${\mathcal M}$ with
probability $p$, and the other with probability $1-p$. The resulting
measurement is given by
\begin{eqnarray}
 {\mathcal Q} & = &   \{ \sqrt{p} \Omega_m^{\mathcal M}  \} \cup
                      \{ \sqrt{1-p} \Omega_n^{\mathcal N} \} \\
              & \equiv &   p {\mathcal M} + (1-p) {\mathcal N} .
\end{eqnarray}
This is a kind of `addition' operation for measurements.

{\em Concatenation:} In this case the observer performs two
measurements, one after the other. If ${\mathcal M}$ is made first
and ${\mathcal N}$ second, then this is equivalent to making the
single measurement ${\mathcal Q}$, where
\begin{eqnarray}
 {\mathcal Q} & = & \{\Omega_n^{\mathcal N}\Omega_m^{\mathcal M}
                       : n = 1,\ldots,N ; m = 1,\ldots,M\} \\
              & \equiv & {\mathcal M} \circ {\mathcal N} .
\end{eqnarray}
One might think of this as a kind of `multiplication' operation for
measurements.

\section{Two Information-Theoretic Capacities}
\label{CCandMS} 
We are interested here in quantifying the amount of information which 
a measurement provides about the system being measured.  In classical 
terms one might think of concepts such as the `accuracy' of a 
measurement, or the `resolving power'.  It is worth noting that, in 
general, the amount of information that the measurement provides 
depends, in a way we will make precise later, on the state of the 
system being measured.  However, in what follows we will be 
specifically interested in quantities which depend only on the 
measurement itself.  Such quantities can be obtained by optimizing 
over the measured state, or simply by fixing it, if a natural choice 
exists.

The first thing to consider when speaking about the information 
obtained in a measurement is what this information is {\em about}.  
For quantum systems, there are two choices, since one can consider the 
information provided about the initial preparation, or the information 
provided about the state which results from measurement.  In classical 
measurements there is only one choice, because these two things are 
the same.

One kind of information, that about the initial preparation, is useful 
in communicating classical information.  If Alice encodes a message in 
a quantum system, then the ability of a measurement to extract 
information about the initial preparation tells us how much classical 
information Bob can obtain from Alice by employing the measurement.  
Thus, a measure of this kind of information is a {\em classical 
capacity}.

Alternatively, one can ask how much information one obtains about the 
final state - that is, the state that we are left with once we have 
made the measurement.  How much one knows about a quantum state can be 
characterized by the von Neumann entropy of this state.  Thus, a 
measure of the amount of information which the measurement extracts 
about a state can be obtained by taking the average decrease in the 
von Neumann Entropy resulting from the measurement.  This motivates 
the definition of the {\em purification capacity} (and the related 
concept of the {\em measurement strength}~\cite{DJJ,FJ}).  We now deal 
with each of these measures of information in turn.

\subsection{Classical Capacity}
\label{CC}

There is now a considerable body of work on the capacity of quantum 
channels for transmitting classical information (see, e.g.  
\cite{Holevo73,SW,Holevo98,Fuchs,3states,Shor,King,HN}).  Here, 
however, we are interested in examining the capacity of a {\em 
measurement} to obtain classical information encoded in a quantum 
state.  In defining the capacity of a channel, one must maximize over 
all measurement strategies and encodings, in order to find the amount 
of information that can be communicated by the channel.  To obtain the 
capacity of a measurement, instead one assumes a perfect channel, 
fixes the measurement and optimizes over all possible encodings, thus 
providing a quantity which characterizes the measurement rather than 
the channel.

In contrast to classical channels~\cite{Shannon}, in considering the 
classical capacity for quantum channels, there is a distinction 
between the `single shot' capacity and the asymptotic capacity.  The 
single shot capacity tells us how much classical information can be 
transmitted by a single use of the channel.  If the channel is a 
single qubit being transmitted from sender to receiver, for example, 
then the single shot capacity tells us how much information can be 
obtained by measuring just the one qubit.  The asymptotic capacity is 
the amount of information that can be transmitted, per qubit, if we 
are allowed to measure multiple qubits, in the limit of an infinity 
number of qubits.  In general, the asymptotic capacity is greater than 
the single shot capacity~\cite{Sup1,Sup2,Sup3}.  However, King and 
Ruskai~\cite{ProdCap} have recently shown that this is only true if 
joint measurements are made across multiple uses of the channel.  
Thus, as a property of a measurement on a single system, the notion of 
classical capacity is well defined without reference to the number of 
systems being measured.

The amount of classical information that is provided by a measurement, 
${\mathcal M}$, about an encoding, $\varepsilon$, and which we will 
denote by $\Delta I_{\mbox{\scriptsize i}}({\mathcal M},\varepsilon)$, 
is the mutual information between the encoding and the output results.  
An encoding is the set of states chosen as the alphabet with which to 
write the information, along with the probability with which each 
state will be selected. We will denote an element of the set of all 
possible encodings by $\varepsilon$, and the set of states and 
probabilities corresponding to this encoding as $\{ \rho_i : i = 
1,\ldots n \}$, and $\{ P(i) : i = 1,\ldots n \}$, respectively.  The 
state of the prepared system from the point of view of the observer 
making the measurement is $\rho = \sum_i P(i) \rho_i$. With these 
definitions, the information provided by the measurement is given by 
\begin{equation}
 \Delta I_{\mbox{\scriptsize i}}({\mathcal M},\varepsilon) =  
                   H[P(i)] - \sum_j Q(j) H[P(i|j)] ,
 \label{forIc}
\end{equation}
where
\begin{equation}
  Q(j) = \mbox{Tr}[\Omega_j^\dagger\Omega_j\rho]
\end{equation}
is the {\em a priori} probability density of the output results,
\begin{equation}
  P(i|j) = \frac{\mbox{Tr}[\Omega_j^\dagger\Omega_j\rho_i] P(i)}
                {\sum_{i}\mbox{Tr}[\Omega_j^\dagger\Omega_j\rho_i] P(i)} 
         = \frac{Q(j|i)P(i)}{\sum_{i}Q(j|i)P(i)}
\end{equation}
is the observer's state of knowledge of the initial preparation on 
receiving the outcome $j$, and $H$ is the entropy.  The quantity 
$Q(j|i)$ introduced above is the probability density of the output 
results given that the initial state is $\rho_i$.  We will continue to 
use this notation throughout; that is, we will denote the probability 
densities of the initial states by $P$, and the densities of outcomes 
by $Q$.  Note that the mutual information is simply the average 
reduction in the entropy of the distribution of the initial 
preparation, $P(i)$, due to the measurement.  It is useful to note 
that the mutual information can also be written in the reverse form 
with input $i$ and output $j$ swapped.  That is
\begin{equation}
 \Delta I_{\mbox{\scriptsize i}}({\mathcal M},\varepsilon) =  H[Q(j)] - \sum_i P(i) H[Q(j|i)] .
   \label{revMut}
\end{equation}
Naturally the classical capacity of a measurement should be defined as 
the supremum of $\Delta I_{\mbox{\scriptsize i}}({\mathcal 
M},\varepsilon)$ over all initial encoding, and is therefore given by
\begin{equation}
 C({\mathcal M}) = \sup_\varepsilon \Delta I_{\mbox{\scriptsize i}} 
                                    ({\mathcal M},\varepsilon) .
 \label{forCap}
\end{equation}

It is natural now to ask how $C$ behaves under the operations of
mixing and concatenation.  For mixing it is clear that
\begin{equation}
 C(p{\mathcal M} + (1-p){\mathcal N}) \leq pC({\mathcal M}) 
                        + (1-p)C({\mathcal N}) .
\end{equation}
This is because, since the observer chooses between measurements 
${\mathcal M}$ and ${\mathcal N}$ on a random basis, the preparer 
cannot know which measurement will be used, and therefore must use the 
same encoding irrespective of the measurement made.  If the optimal 
encoding for both measurements is the same, then we have equality.  
Otherwise one measurement necessarily extracts less information than 
is optimal, hence the inequality.

The behavior under concatenation is straightforward to obtain for 
classical measurements.  If we denote the mutual information between 
two sets of random variables $X$ and $Y$ as $M(X,Y)$, then the 
capacity of the concatenated measurement is the supremum of $M(i|j,k)$ 
over initial encodings, where $i$ labels the elements of the encoding, 
and $j$ and $k$ label the outcomes for the respective measurements 
${\mathcal M}$ and ${\mathcal N}$.  Rearranging the expression for 
$M(i|j,k)$, one finds that
\begin{equation}
  M(i|j,k) = M(i|j) + M(i|k) - M(j|k) .
\end{equation}
The point here is that, while two measurements may be independent, the 
outcomes of the measurements, when made in sequence on the same 
system, are not, in general, uncorrelated.  Since $M(j|k) \geq 0$, we 
have $M(i|j,k) \leq M(i|j) + M(i|k)$.  Since the classical capacity of 
${\mathcal M}$ and ${\mathcal N}$ are the respective supremums of 
$M(i|j)$ and $M(i|k)$ over initial encodings, we have
\begin{eqnarray}
 C({\mathcal M} \circ {\mathcal N}) & = & \sup_{\varepsilon} M(i|j,k) \nonumber \\
           & \leq & \sup_{\varepsilon'}  [ M(i|j) + M(i|k) ] \nonumber \\
           & \leq & \sup_{\varepsilon'}   M(i|j) + \sup_{\varepsilon''} M(i|k) ] \nonumber \\
           & = & C({\mathcal M}) + C({\mathcal N}) .
   \label{ccsubadd}
\end{eqnarray}
The first inequality is only saturated when $M(j|k) = 0$ and the
second is only saturated when the optimal encoding is the same for
both measurements.

For quantum measurements at first it is tempting to suggest that $C$ 
might also be subadditive.  However, if we allow all classes of 
measurements, in particular measurements in which the operators 
$\Omega_n$ are not purely positive, but have a unitary component, then 
it is not hard to find examples which break the inequality.  That is,
\begin{equation}
 \exists \; {\cal M},{\cal N} \;\; , \;\;\; C({\mathcal M} \circ {\mathcal N}) 
                     > C({\mathcal M}) + C({\mathcal N}) .
\end{equation}
Nevertheless, it is important to point out that the same can be said 
for the mutual information for classical operations if we allow 
conditional deterministic transformations.  As a result we are 
prepared to conjecture that for {\em bare} measurements the 
purification capacity satisfies the classical relation, 
Eq.(\ref{ccsubadd}).

\subsection{Purification Capacity}
\label{MS} While the classical capacity was motivated by considering 
the transmission of classical information, the {\em purification 
capacity} is motivated by considering the problem of state preparation 
and more generally quantum feedback control~\cite{DJJ,FJ}.  In this 
case, instead of being interested in the amount of information one 
obtains about the {\em initial} preparation, one is interested in how 
well one knows the {\em final} state that results from the 
measurement.  This is because the more pure the final state, the more 
one can subsequently control the system.  That is, one is interested 
in the amount by which, on average, the measurement purifies the state 
of the system.

For the purposes of control, the von Neumann entropy is a sensible
measure of the observer's uncertainty of a state, since it measures
the minimum possible entropy of the outcomes of a future measurement.
That is, if the observer has the ability to perform any
measurement, the von Neumann entropy measures the unavoidable
residual unpredictability of the measurement result. The outcome
of a measurement on a state with zero von Neumann entropy can, in
the best case, be predicted perfectly.

To characterize the amount of information that a measurement
provides about the final state one can therefore average the von
Neumann entropies of all the possible final states over the
measurement outcomes. Subtracting this from the von Neumann
entropy of the initial state gives us a measure of the increase in
the observer's information about the state of the system provided
by the measurement:
\begin{equation}
  \Delta I_{f} ({\mathcal M},\rho) = S(\rho) - \sum_i P(i) S(\rho_i) ,
\end{equation}
where ${\mathcal M}$ is the measurement and $\rho$ is the initial
state. Nielsen has shown that, as one would intuitively expect, $\Delta 
I_{f}$ is non-negative~\cite{Nielsen}. (See also~\cite{FJ} for a 
remarkably simple proof of the this result discovered by Fuchs.)

In order to obtain a quantity which depends only on the measurement, 
two possibilities are to fix the initial state, or to maximize $\Delta 
I_{f}$ over all possible initial states.  If we choose the second 
option, then we have a quantity which is the final-state equivalent of 
the classical capacity. We will refer to this as the {\em 
purification capacity}, and denote it by $K$. Thus we define
\begin{equation}
  K ({\mathcal M}) = \sup_{\rho} \Delta I_{f} ({\mathcal M},\rho) .
\end{equation}
If instead we wish to fix the initial state, which was the procedure 
adopted in~\cite{DJJ}, then it is important to choose a state which is 
invariant under unitary transformations, so that the resulting 
quantity, which was referred to in~\cite{DJJ} as the {\em measurement 
strength}, satisfies the intuitive notion that it should be invariant 
under unitary transformations of the measurement.  In 
Refs.~\cite{DJJ,FJ} the measurement strength, which we will denote by 
$K_{I}$, was defined by fixing the initial state to be the maximal 
uncertainty state $\rho = I/N$. This gives
\begin{equation}
  K_{I} ({\mathcal M}) = \ln(N) - \sum_n P_n S[\tilde{E_n}] ,
\end{equation}
where $\tilde{E_n} = E_n/\mbox{Tr}[E_n]$, and we have used the fact 
that the eigenvalues of $\Omega_n\Omega_n^\dagger$ are the same as 
those of $E_n$~\cite{MO,AndoMaj}.  By the definition of $K$ we have 
$K_{I}({\cal M}) \leq K({\cal M})$.

It is now natural to ask how the purification capacity and the 
measurement strength behave under the operations of mixing and 
concatenation.  Under mixing one has trivially that $K_{I}$ is linear, 
ie.
\begin{equation}
  K_{I}(p{\mathcal M} + (1-p){\mathcal N}) = pK_{I}({\mathcal M}) + 
                                            (1-p)K_{I}({\mathcal N}) ,
\end{equation}
and that $K$ is concave:
\begin{equation}
  K(p{\mathcal M} + (1-p){\mathcal N}) \leq pK_{I}({\mathcal M}) + 
                                            (1-p)K_{I}({\mathcal N}) .
\end{equation}
While it is not immediately obvious, for bare measurements $K_{I}$ is 
invariant under a change in the order in which measurements are 
concatenated.  That is,
\begin{equation}
  K_{I}({\mathcal M} \circ {\mathcal N}) = K_{I}({\mathcal N} \circ {\mathcal 
  M}) \;\; , \;\;\; {\mathcal M},{\mathcal N} \in {\mathcal B} ,
\end{equation}
where ${\mathcal B}$ denotes the set of all bare measurements.
This follows almost immediately from the fact that, for any two 
operators $A$ and $B$, $AB$ has the same eigenvalues as 
$BA$~\cite{MO,AndoMaj}. 

For general quantum measurements, it is not hard to show that neither 
the measurement strength nor the purification capacity satisfy the 
classical relation, given by Eq.(\ref{ccsubadd}) (Note that for classical 
measurements $K=C$.) However, we conjecture that for bare 
measurements both the strength and purification capacity satisfy the 
classical relation, e.g.
\begin{equation}
  K({\mathcal M} \circ {\mathcal N}) \leq  K({\mathcal M}) + K({\mathcal N}) 
                 \;\; , \;\;\; {\mathcal M},{\mathcal N} \in {\mathcal 
                 B} .
\end{equation}

\section{Two Measures of Disturbance}
\label{Dist}

Although disturbance is not the primary focus of this article, we feel 
that it is worth pointing out here that, corresponding to the two 
measures of information discussed in the previous section, there are 
two natural information-theoretic measures of the disturbance caused 
by a measurement.  The first, which we might refer to as disturbance 
to the input, measures the amount by which classical information 
encoded in a quantum state is degraded by a measurement.  That is, it 
tells how much less classical information we can extract from the 
system about the initial preparation after having made the 
measurement.  For this purpose we will assume that the observer who 
wishes to extract the information does not have access to the results 
of the disturbing measurement.  The reason for using this definition 
is that it results in a sharp contrast between classical and quantum 
measurements; for classical measurements this disturbance is zero, but 
can be non-zero for quantum measurements.

The second kind of disturbance, which we might refer to as disturbance 
to the output, is the difference between the entropy of the initial 
state, and the entropy of the state given by averaging all the final 
states.  This measures the amount of noise that the measurement is 
feeding into the system.  This is understood most easily by thinking 
of a noise-driven classical system.  The random changes that the 
quantum system experiences as a result of the random outcomes of the 
measurement are like the random kicks experienced by a classical 
system under the influence of a noisy force.  The strength of this 
noise can be characterized by the entropy increase of the probability 
distribution for the state of the system given that the observer has 
no knowledge of which kick will occur.  This quantum measure of 
disturbance is simply the equivalent calculation for a quantum system.  
This kind of disturbance is relevant for quantum feedback control, and 
was considered in~\cite{DJJ,FJ}.  The disturbance caused by a 
classical measurement under this definition is zero, whereas, for bare 
quantum measurements, it is greater than or equal to 
zero~\cite{AndoMaj}.  Thus, both measures provide a precise concept of 
the general notion that quantum measurements can produce a 
disturbance~\cite{folkL}, whereas classical measurements do not.

Note that the definitions of disturbance which have been most studied 
in the literature to date, for example in the work by 
Fuchs~\cite{FuchsDist}, Banaszek~\cite{Banaszek} and 
Barnum~\cite{Barnum}, characterize disturbance to the input. It is 
therefore our first measure which is most closely related to these 
notions of disturbance.

\subsection{Disturbance to the Input}

Given an encoding, $\varepsilon$, the optimal amount of information 
which an observer Alice can obtain about a message encoded by Bob 
using that encoding is called the {\em accessible 
information}~\cite{AccessInfo}.  If a third observer, Eve, makes a 
measurement, ${\mathcal M}$, on the system in which the information is 
encoded, and does not relay the result of the measurement to the 
Alice, then from the point of view of the first observer the state of 
the system is transformed by
\begin{equation}
   \rho \rightarrow \rho' = \sum_n \Omega_n \rho \Omega_n^\dagger ,
\end{equation}
where the $\Omega_n$ are the measurement operators for ${\mathcal M}$.  
As a result, the encoding is transformed from $\varepsilon$ to 
$\varepsilon'$ where
\begin{equation}
   \varepsilon' = \{ p_i, \sum_n \Omega_n \rho_i \Omega_n^\dagger : i = 1,\cdots, N\} .
\end{equation}
The disturbance caused by the measurement ${\mathcal M}$ to the 
ensemble $\varepsilon$ can then be characterized by the difference 
between the accessible information of $\varepsilon$ and that of 
$\varepsilon'$, which we can write as
\begin{equation}
   D_{\mbox{\scriptsize i}}(\varepsilon) = I_{\mbox{\scriptsize access}}(\varepsilon) - I_{\mbox{\scriptsize access}}(\varepsilon') .
\end{equation}
To obtain a measure of disturbance which characterizes the measurement 
alone, we can do one of at least two things.  The first is simply to choose 
$\varepsilon$ to the be the optimal encoding for a perfect channel, 
which, for an $N$-dimensional system, is an $N$- dimensional 
orthonormal basis.  In this case the first term is simply the capacity 
of the perfect $N$-dimensional quantum channel.  The second term, 
however, is not the maximal amount of information which can be 
transmitted from Alice to Bob given that Eve makes the measurement 
${\mathcal M}$, since the optimal encoding $\varepsilon$ for the 
perfect channel, may not be the optimal encoding for the channel 
described by the measurement operators $\Omega_n$.  Thus, a second 
approach to characterizing the disturbance of the measurement is to 
choose the second term in the expression for $D_{\mbox{\scriptsize 
i}}(\varepsilon)$ to be the accessible information when $\varepsilon'$ 
is such as to make this optimal.  The result is simply the single shot 
capacity of the channel described by the measurement operators for 
${\mathcal M}$.  Thus, the input disturbance is essentially a channel 
capacity.  If we write the capacity of the perfect channel as 
$C_{\mbox{\scriptsize cap}}({\mathcal I})$, and that of the channel 
described by ${\mathcal M}$ as $C_{\mbox{\scriptsize cap}}({\mathcal 
M})$, then the input disturbance of ${\mathcal M}$ is
\begin{equation}
   D_{\mbox{\scriptsize i}} = C_{\mbox{\scriptsize cap}}({\mathcal I}) - C_{\mbox{\scriptsize cap}}({\mathcal M}) .
\end{equation}
For classical measurements on classical systems $D_{\mbox{\scriptsize 
i}}=0$.  For quantum measurements $D_{\mbox{\scriptsize i}} \geq 0$.  
To see that $D_{\mbox{\scriptsize i}}$ is non-negative one merely 
needs to note that the accessible information is a quantity which is 
maximized over all measurements.  As a result, the measurement 
${\mathcal M}$ made by Eve can be included as part of Alice's possible 
strategies for obtaining information in the case of the perfect 
channel.  Thus, the case in which Eve makes her measurement is just a 
special case of the perfect channel in which Alice, as part of her 
strategy, first makes Eve's measurement and then throws away the 
information before making any further measurements.  Therefore, the 
capacity of the channel described by ${\mathcal M}$ cannot be more 
than that of the perfect channel, and hence the disturbance 
$D_{\mbox{\scriptsize i}}$ is a non-negative quantity.

\subsection{Disturbance to the Output}

The above measure of disturbance is concerned with the reduction in 
the ability of an observer to obtain information about an initial 
preparation, given that another observer has made a measurement on the 
system to which the first observer has no access.  Now we consider the 
reduction in an observer's knowledge of the {\em final} state of the 
system, given that a second observer has made a measurement to which 
the first has no access.  If the initial state is $\rho$, then the 
final state of the system after a measurement ${\mathcal M}$, from the 
point of view of the first observer, is
\begin{equation}
   \rho' = \sum_n \Omega_n  \rho \Omega_n^\dagger .
\end{equation}
The reduction in the observer's knowledge of the state of the system, 
$\rho$, caused by ${\mathcal M}$, may be characterized by the increase 
in the von Neumann entropy of $\rho'$ over $\rho$.  That is,
\begin{equation}
   D_{\mbox{\scriptsize o}}(\rho) = S[\rho'] - S[\rho] .
\end{equation}
This quantity is zero for classical measurements, and Ando has shown 
that it is non-negative for all bare measurements~\cite{AndoMaj}.

No doubt there is more than one way to use the above measure to obtain 
a quantity which characterizes the measurement alone, and it may well 
be that different choices may be motivated by different applications.  
Nevertheless, one reasonable procedure is to minimize 
$D_{\mbox{\scriptsize o}}(\rho)$ over all initial states.  That is, to 
define the output disturbance of a measurement as
\begin{equation}
   D_{\mbox{\scriptsize o}} \equiv \inf_\rho S \left[ \sum_n \Omega_n  \rho \Omega_n^\dagger \right] - S[\rho] .
\end{equation}
This measures the least disturbance that can be achieved with the 
given measurement if one is free to select the initial state.  Under 
this definition one has the desirable result that all commutative
measurements have $D_{\mbox{\scriptsize o}} = 0$.

\section{Information Gathering and State-Space Symmetry}
%\section{Classical Capacity and State-Space Symmetry}
\label{IGSSS}

\subsection{Classical Measurements}
Since classical measurements have already been introduced in 
Section~\ref{CandQM}, we will merely pause to clarify our definitions: 
By a {\em classical} measurement, we will mean a measurement which can 
be performed on a classical system.  Thus a classical measurement is 
not only one in which all the measurement operators commute, but also 
one in which is it understood that the density operator is also 
required to be diagonal in the same basis as these operators.  We will 
refer to measurements in which the only restriction is that the 
operators commute as {\em commutative} quantum measurements.  The 
reason for this distinction is that it is possible to enhance the 
properties of measurements with unitary transformations~\cite{rapidP}.  
Since such transformations require transforming the density matrix so 
that it is diagonal in other bases, one must make a distinction 
between classical measurements and commutative measurements; the 
latter can be enhanced in ways in which the former cannot.

For classical measurements, there is no difference between the 
information about the initial encoding, and that about the state which 
exists after the measurement, because the measurement does not 
interfere with the initial preparation; both $\Delta 
I_{\mbox{\scriptsize i}}({\mathcal M},\varepsilon)$ and $\Delta I_{f} 
({\mathcal M},\rho)$ reduce to the classical mutual information, which 
we will write simply as $\Delta I({\mathcal M},\rho)$.  Thus 
$K = C$, and the measurement strength $K_{I}$ is 
the mutual information between the output results and the initial 
preparation when the encoding is the uniform ensemble over the all the 
available initial states.

% now say that this implies that the average information gain decreases as information is obtained.
% Or do we say this in the section on Classical capacity?

Now, the mutual information is the average decrease in the Shannon 
entropy of the observer's state of knowledge as a result of the 
measurement, and depends, in general upon not only the measurement but 
also the initial state.  How exactly, does it depend on the initial 
state?  Since the entropy of the state cannot decrease below zero, one 
might suggest that the information gain in the measurement decreases 
as the uncertainty in the observer's initial state of knowledge 
decreases: the more you know the less you find out.  However, from the 
fact that the mutual information is not always maximized when the 
ensemble has maximal entropy, we know that it is possible for the 
reverse to be true for at least some measurements and some states.  So 
is there a sense in which a reduction in information gain with 
increasing knowledge is a fundamental property of measurement? 

To answer this question, we consider a class of measurements which we 
will refer to as {\em permutation-symmetric} measurements.  We define 
these measurements as those which are symmetric under the permutation 
of two classical basis states.  That is, when any two states 
$|i\rangle$ and $|j\rangle$ are swapped, the set of measurement 
operators remains unchanged --- the various measurement operators 
merely interchange among themselves.  The motivation for such a 
definition is the observation that the information extraction 
capability (or accuracy, or resolution) of a measurement can, in 
general, vary across the state space, and it is this that causes some 
measurements to provide a larger average amount of information when 
applied to states with less that maximal entropy.  The definition of 
permutation-symmetric measurements ensures that they have the same 
resolving power over all of state space.  We now show that these 
measurements have the property which we seek.  First it is helpful to 
introduce some notation.  We will denote the vector of eigenvalues of 
a matrix $A$ by ${\bf \lambda}(A)$.  If we write $\rho \prec \sigma$ 
then we will mean that ${\bf \lambda}(\rho) \prec {\bf 
\lambda}(\sigma)$~\cite{Majnote}. 

\vspace{1mm} {\em Theorem:} If ${\mathcal M}$ is classical and
permutation-symmetric, then for any two classical states $\rho$
and $\sigma$, $\rho \prec \sigma$ implies that $\Delta I({\mathcal
M},\rho) > \Delta I({\mathcal M},\sigma)$. (Note that a more concise way of 
saying this is that $\Delta I({\mathcal M},\rho)$ is {\em 
Schur-concave} in $\rho$~\cite{MO}).

\vspace{1mm} {\em Proof:} To begin we note that given a diagonal 
operator $\Omega$ of dimension $N$, the sum of the $N!$ operators, 
$\{\Omega_m\}$ which are the different permutations of $\Omega$, is 
proportional to the identity.  This means that the operators $\{\alpha 
\Omega_m\}$ form a valid measurement for some $\alpha$.  In what 
follows when we refer to an operator $\Omega$ which is used to 
generate a permutation-symmetric measurement by taking all its 
permutations, we will always define $\Omega$ so that 
$\mbox{Tr}[E]=\mbox{Tr}[\Omega^{2}]=1$, in which case 
$\alpha=1/\sqrt{(N-1)!}$.  We will refer to a measurement generated by 
a single operator $\Omega$ as an irreducible permutation-symmetric 
measurement (IPM).  Clearly all permutation-symmetric measurements can 
be written as mixtures (in the sense of the mixing operation described 
in section~\ref{MOps}) of IPM's.  Now consider the mutual information 
for an IPM. This may be written as
\begin{equation}
  \Delta I ({\mathcal M},\rho) = H[Q_\rho(m)] - \sum_n P_\rho(n)
  H(Q_\rho(m|n)),
\end{equation}
where $Q_\rho(m)$ is the distribution of the measurement outcomes, 
$P_\rho(n)$ is the distribution of the initial states (being the 
diagonal of $\rho$), and $Q_\rho(m|n)$ is the distribution of 
measurement outcomes given that the initial state is $n$.  The first 
thing to note is that due to the permutation symmetry, 
$H(Q_\rho(m|n))$ is the same for all $m$.  Thus, the second term is 
independent of the initial state $\rho$, and we need merely show that
\begin{equation}
  H[Q_\rho(n)] > H[Q_\sigma(n)] \;\;\; \mbox{when} \;\;\; \rho \prec \sigma .
\end{equation}
For this it is sufficient to show that $Q_\rho \prec Q_\sigma$ if 
$\rho \prec \sigma$.  That is, that the operation that transforms 
$\rho$ to $Q_\rho$ preserves majorization.  The operation that 
transforms $\rho$ to $Q_\rho$ may be written as
\begin{equation}
  Q_\rho = A {\bf \rho}_v ,
\end{equation}
where ${\bf \rho}_v$ is the vector consisting of the diagonal of 
$\rho$, and $A$ is a matrix whose rows are the diagonals of the 
operators $\Omega_m\Omega_m^\dagger$.  The rows of $A$ therefore 
consist of all the permutations of any given row.  At this point we 
can invoke a theorem of Chong~\cite{Chong74,MO}, who has shown that 
the operation of multiplication by a matrix preserves majorization if 
the matrix is such that the rows form a permutation invariant set.  
That is, if all permutations of any row of the matrix are also rows of 
the matrix.  We will refer to matrices of this form in what follows as 
{\em Chong matrices}.  This proves the result for IPM's.  Since every 
permutation-symmetric measurement is a mixture of IPM's, and $\Delta 
I$ is linear under mixing, the result holds for all 
permutation-symmetric measurements.  $\square$

That is, once we have eliminated the state-space dependence of the 
resolution of measurements, a fundamental property remains.  This is 
that, the more one knows, the less is the amount that one will learn 
by applying a given measurement.

As a corollary of this, we have that the classical capacity of a 
classical irreducible permutation-symmetric measurement is attained by 
the uniform distribution, and is therefore given by
\begin{equation}
  C = \ln N - S(E) ,
  \label{capCIPM}
\end{equation}
where $S$ is the von Neumann entropy, and $E\equiv\Omega^{2}$.  In 
addition, it is worth noting that since both classical channels and 
classical measurements are defined by a complete set of conditional 
probabilities, classical measurements and classical channels are one 
and the same thing.  Thus, Eq.(\ref{capCIPM}) also gives the capacity 
of irreducible permutation-symmetric classical channels.  The capacity 
of a general permutation-symmetric measurement 
\begin{equation}
  {\mathcal M} = \sum_{n=1}^{N} p_{n} {\mathcal N}_{n},
  \label{genPM}
\end{equation}
where each of the measurements ${\mathcal N}_{n}$ is generated by 
the operator $E_{n}$ is thus 
\begin{equation}
  C = \ln N - \sum_{n=1}^{N} p_{n} S(E_{n}) .
  \label{capCPM}
\end{equation}

\subsection{Unitarily Covariant Measurements}

In the previous subsection we found a class of classical measurements 
for which the information provided by the measurement always increased 
with the initial uncertainty.  In this section we turn our attention 
to quantum measurements.  In this case, it is the Unitarily Covariant 
Measurements (UCM's), to be defined below, which are the equivalent of 
the classical permutation-symmetric measurements, in that they are 
invariant under all reversible transformations of the state-space.  
However, in considering the information gathering properties of UCM's 
we need to be a little more precise about what we mean by `information 
about the initial state'.

In the previous section we assumed that the information was always 
encoded in individual (or pure) states, rather than probability 
densities of states (or mixtures).  What this assumption meant is that 
$\Delta I_{i}$, which measures the information provided about the 
encoded message, {\em also} measured the information obtained about 
which state the system was initially in.  Note that, since classical 
systems are always in {\em some} `pure' state, it is not ambiguous to 
ask what state the system is in, even if the initial encoding is in 
mixed states.  However, if the initial encoding {\em does} use mixed 
states, $\Delta I_{i}$ no longer measures the information that the 
measurement provides about the what state the system is in.

When we consider obtaining a quantum equivalent of the theorem of the 
previous section, then we need to use the equivalent notion of 
information - that is, information about what {\em pure} state the 
system is in.  Now, in quantum mechanics, in general the system does 
not have to be in {\em any} pure state, since an alternative exists - 
it could be entangled with another system.  Thus, we need to be clear 
that the situation we are concerned with here is that in which the 
system is known to be in {\em some} pure state.  The information we 
are concerned with is the information provided about what that state 
is, and this is what $\Delta I_{i}({\mathcal M},\varepsilon)$ measures 
so long as $\varepsilon$ is an ensemble of pure states. In what 
follows our analysis will therefore be restricted to ensembles of pure 
states.

 % We will now show	that the UCM's are a class of quantum measurements 
 % for which the information gain about	the	initial	pure state ensemble, $\Delta 
 % I_{\mbox{\scriptsize	i}}({\mathcal M},\varepsilon)$,	always increases 
 % with	the	initial	uncertainty

To begin, the unitarily covariant measurements~\cite{Barnum, Ucov} are 
defined as being those measurements which are invariant under every 
unitary transformation of the measurement operators; that is, when all 
the measurement operators are transformed by a given unitary, all the 
measurement operators merely transform among themselves.  Thus, every 
UCM has a continuum of measurement operators, labeled by the unitary 
transforms $U$.

Now, we wish to show that the UCM's are a class of quantum 
measurements for which the information gain about the initial pure 
state ensemble, $\Delta I_{\mbox{\scriptsize i}}({\mathcal 
M},\varepsilon)$, always increases with the initial uncertainty.  Now 
we must ask, `the initial uncertainty of what?'.  Since, in general, 
$\Delta I_{\mbox{\scriptsize i}}({\mathcal M}, \varepsilon)$ is a 
function of the initial ensemble, one might expect that we would have 
to consider some uncertainty property of the ensemble, rather than the 
initial state, $\rho$.  However, it turns out that this is not the 
case:

\vspace{1mm} {\em Theorem:} The information $\Delta 
I_{\mbox{\scriptsize i}}({\mathcal M},\varepsilon)$, obtained about a 
pure-state ensemble $\varepsilon$ by a unitarily covariant measurement 
${\cal M}$, depends on the ensemble only through the density matrix 
$\rho=\sum_{i}P(i)|\psi_{i}\rangle\langle \psi_{i}|$ where 
$\varepsilon = \{P(i),|\psi_{i}\rangle\}$.

\vspace{1mm} {\em Proof:} The expression for the information gain is 
\begin{eqnarray}
	 \Delta I ({\mathcal M},\varepsilon) & = & H[Q_\rho(U)] \nonumber \\ 
                                  & - & \sum_m P_\rho(|\psi_{m}\rangle) 
                                        H(Q_\rho(U||\psi_{m}\rangle)) .
\end{eqnarray}
Since ${\cal M}$ is unitarily covariant, 
$H(Q_\rho(U||\psi_{m}\rangle))$ is the same for all initial states 
$|\psi_{m}\rangle$, and therefore the second term is the same for all 
initial ensembles.  As a result, $\Delta I ({\mathcal M},\rho)$ 
depends only on the first term, which depends only on $\rho$.  
$\square$

In view of the above result, we now show that 
under unitarily covariant measurements~\cite{Ucov}, $\Delta 
I_{\mbox{\scriptsize i}}({\mathcal M},\rho(\varepsilon))$ is Shur-concave 
in $\rho$.

\vspace{1mm} {\em Theorem:} If ${\mathcal M}$ is unitarily covariant, 
then for any two states $\rho$ and $\sigma$, $\rho \prec \sigma$ 
implies that $\Delta I_i({\mathcal M},\rho(\varepsilon)) > \Delta 
I_i({\mathcal M},\sigma(\varepsilon'))$, where $\varepsilon$ and 
$\varepsilon'$ are ensembles of pure states.

{\em Proof:} First we note that given a positive operator $\Omega$, we 
have
\begin{equation}
  \int U E U^\dagger d\mu(U) = \mbox{Tr}[E] I ,
  \label{covPOVM}
\end{equation}
where $E=\Omega^{2}$ and $d\mu(U)$ is the (unitarily covariant) Haar 
measure over unitary operators $U$~\cite{Ucov}.  Thus, to generate a 
covariant measurement from an operator $\Omega$, we first scale 
$\Omega$ so that $\mbox{Tr}[E]=1$.  We will refer to a UC measurement 
generated from a single operator $\Omega$ as an irreducible UC 
measurement.  Now, Eq.(\ref{covPOVM}) also tells us that for any 
covariant POVM, each subset of the measurement operators which is 
closed under unitary transformations form themselves a POVM. As a 
result we can restrict ourselves to measurements in which all the 
measurement operators are obtained from each other by a unitary 
transform; all other unitarily covariant measurements can be obtained 
from measurements of that form by mixing.

Now consider the information gain
\begin{eqnarray}
 \Delta I ({\mathcal M},\rho) & = & H[Q_\rho(U)_{}] \nonumber \\
                               & - & \sum_m P_\rho(|\psi_{m}\rangle) 
                                     H(Q_\rho(U||\psi_{m}\rangle)) .
\end{eqnarray}
Now, since the measurement is unitarily covariant, 
$H(Q_\rho(U||\psi_{m}\rangle))$ is independent of the state 
$|\psi_{m}\rangle$.  As a result, the second term in the expression 
for $\Delta I$ is independent of the ensemble probabilities 
$P_{\rho}(|\psi_{m}\rangle)$, and therefore of the initial state 
$\rho$.  Thus we need merely consider the first term, 
$H[Q_{\rho}(U)]$.  Now since
\begin{eqnarray}
	H[Q_{\rho}(U)] & = & \int f(\mbox{Tr}[U E U^\dagger\rho]) d\mu(U)
\end{eqnarray}
(where $f(x) = - x\ln x$), $H$ is invariant under a unitary 
transformation of $\rho$, and as a result we can always choose the 
eigenvectors of $\rho$ to be the same as the eigenvectors of $E$.  
Denoting these eigenvectors by $\{|j\rangle\}$, and the eigenvalues of 
$\rho$ and $E$ as $\{\lambda_{j}\}$ and $\{E_{i}\}$, respectively, we 
have
\begin{eqnarray}
	Q_{\rho}(U) & = & \mbox{Tr}[UEU^{\dagger}(\sum_{j}\lambda_{j}
	                  |j\rangle\langle j|)] 
	\nonumber \\
	            & = & \sum_{j} \left[ \sum_{i} E_{i} M_{ij}(U) \right] 
	                  \lambda_{j} ,
\end{eqnarray}
where $M_{ij}(U) \equiv |U_{ij}|^{2}$ is a doubly stochastic matrix.  

The quantity in the square brackets has two indices: the column 
index, $j$, is discrete and the `row' index, $U$, is continuous.  The 
sum over $j$ is a discrete-to-continuous version of matrix 
multiplication.  This continuous nature will make the following 
discussion more involved than that in previous section, but the 
complication is primarily technical.

First we make the problem discrete by choosing a finite set of points 
on which to sample the continuous set of unitary transformations $U$.  
Let us denote the set of points as $\{U_{k} : k=1,\ldots,L\}$, and the 
sampled values of $Q$ at those points, as $Q_{\rho}(U_{k})$.  The set 
of these values is now a vector with $L$ elements.  The discrete 
transformation may now be written as
\begin{eqnarray}
	Q_{\rho}(U_{k}) & = & \sum_{j} \left[ \sum_{i} E_{i} M_{ij}(U_{k}) 
                     	           \right] \lambda_{j} .
\end{eqnarray}
Now, for every sample point $U_{k}$, we have a doubly stochastic matrix, 
$M_{ij}(U_{k})$.  The fact that the continuum transformation includes  
every value of $U$ (that is, every unitary transformation), means 
that, given any sampling $\{U_{k} : k=1,\ldots,L\}$, we can extend 
this sampling to include every column permutation of every one of the 
matrices $M_{ij}(U_{k})$.  (By a column permutation we mean an 
operation in which one or more of the columns of $M$ are swapped.)  
That is, we can choose a new set, $\{U_{l}:\}$, such that it contains 
$\{U_{k} : k=1,\ldots,L\}$ as a subset, and the set 
$\{M_{ij}(U_{l}):l=1,\ldots,L'\}$ is such that for every element $M$, 
every column permutation is also an element.  The transformation for 
this extended set is now majorization preserving, since
\begin{eqnarray}
	Q_{\rho}(U_{l}) & = & \sum_{j} \left[ \sum_{i} E_{i} M_{ij}(U_{l}) 
                       	           \right] \lambda_{j} \\
	                & = & \sum_{j} \tilde{M}_{lj} \lambda_{j} ,
\end{eqnarray}
and $\tilde{M}_{lj}$ is a Chong matrix.  This is because the 
operation of summing over the index $i$ in the first line above simply 
involves summing each of the columns of each stochastic matrix 
$M_{ij}$, to create one row of the matrix $\tilde{M}_{lj}$.  Since 
every column permutation of each $M_{ij}$ exists, the resulting 
matrix, $\tilde{M}_{lj}$, is such that every permutation of each row 
of $\tilde{M}_{lj}$ appears as another row of $\tilde{M}_{lj}$, which 
is Chong's condition.

Our goal is naturally to take the limit as $L$ tends to infinity, to 
obtain the continuum result that we want. However, before we do so we 
must take into account the measure. In calculating the entropy of the 
density $Q_{\rho}(U)$, we integrate over the Haar measure. A
discrete approximation of this integral for our vector 
$Q_{\rho}(U_{l})$ is
\begin{eqnarray}
	H[Q_{\rho}(U_{l})] = \sum_{l=1}^{L'} \Delta(U_{l}) 
	                     f[Q_{\rho}(U_{l})] ,
\end{eqnarray}
where as before $f(x) = -x\log x$, and $\Delta(U_{l})$ is the Haar 
volume associated with each sample point $U_{l}$.  It is important for 
our purposes that in the above equation we associate the same Haar 
volume with all sample points which are obtained from each other by a 
permutation.  (We will refer to the $L$ sets containing $N!$ points, 
all permutations of each other, as the $L$ permutation-invariant 
sets.)  To see that this is possible, for any value of $L$, one first 
notes that the group of permutations forms a finite subgroup 
(containing $N!$ elements) of the group of all unitary 
transformations.  As such, it can be used to divide the set of all 
unitaries up into $N!$ subsets, where each is the image of the others 
under the action of a permutation.  Since the Haar measure is 
invariant under unitary transformations, each of these subsets has not 
only the same Haar volume, but any region defined within one has the 
same volume when that region is mapped to another.  Because of this 
last property, we can do the following: To ensure that all the points 
within a given permutation-invariant set have the same associated Haar 
volume, we can (for example) chose our original $L$ points all to lie 
within one of the $N!$ factorial subsets, choosing the Haar volumes 
associated with each in whatever way we see fit.  When we extend the 
set of points to include all permutations, the regions associated with 
the points in the first subset are mapped to each of the other $N!-1$ 
subsets.  Each of the new points thus has the same Haar volume as the 
original point of which it is the image, which means that all the 
points in a given permutation-invariant set have an associated region 
with the same Haar volume.

Thus, we can break the vector of points $U_{l}$ into $L$ subvectors, 
where the points in each are the points of one of the $L$ 
permutation-invariant sets, and which consequently all have the same 
associated Haar volume.  The importance of this is that each of these 
subvectors is obtained from the $\lambda_{j}$'s by a Chong matrix.  We 
can now write the above summation as,
\begin{eqnarray}
	H[Q_{\rho}(U_{l})] = \sum_{n=1}^{L} \Delta(U_{n}) \left[ \sum_{l_{n}} 
	f[Q_{\rho}(U_{l_{n}})] \right] ,
\end{eqnarray}
which is simply a weighted sum of the entropies of each of the 
subvectors. Since each of the subvectors is obtained from the vector 
$\lambda(\rho)$ by a transformation of Chong form, the entropy of each is 
Shur-concave in $\lambda(\rho)$, and hence 
\begin{eqnarray}
	H[Q_{\rho}(U_{l})] > H[Q_{\sigma}(U_{l})] 
	\label{HH}
\end{eqnarray}
if $\rho\prec\sigma$. Now, finally, we can take the limit 
$L\rightarrow\infty$, so that we recover the Haar integral, and 
$H[Q_{\rho}(U_{l})] \rightarrow H[Q_{\rho}(U)]$. Since we know that 
Eq.(\ref{HH}) is true for each $L$ in the sequence, it is true in the 
limit, and we obtain our result for all ICM's. Since all unitarily 
covariant measurements can be obtained ICM's by mixing (i.e. 
averaging), the result follows for all unitarily covariant measurements. 
$\Box$ 

The following corollary is an immediate consequence:

{\em Corollary:} The classical capacity of a unitarily 
covariant measurement is obtained by the uniform ensemble over any 
basis.

In addition to the above result, we conjecture that the final 
information, $\Delta I_{f} ({\mathcal M},\rho)$, is also Shur-concave 
in $\rho$ for all unitarily covariant measurements. 

\subsection{Classical Capacities for Symmetric and Covariant Measurements}
To obtain the classical capacity of a quantum measurement ${\cal M}$ 
one must optimize $\Delta I_{\mbox{\scriptsize i}}($M$,\varepsilon)$ 
over all initial encodings.  However, it turns out the complexity of 
this procedure is significantly reduced for commutative measurements.  
This is because, as we now show, the classical capacity of a 
commutative quantum measurement is the same as that of the equivalent 
classical measurement.  (A commutative and a classical measurement will 
be said to be equivalent when the measurement operators of the 
commutative measurement are, upon diagonalization, the same as those of 
the classical measurement.)  This gives us the classical capacity of 
all Commutative Permutation-Symmetric (CPS) measurements in terms of the 
previous results for classical permutation-symmetric measurements.

 % While the PSC measurements are simply the classical permutation
 % symmetric measurements when made	on quantum systems,	it is
 % probably	worth reviewing	their definition here. Since all the
 % measurement operators of	${\cal M}$ commute,	they may all be
 % diagonalized	in the same	basis. We can thus think of	the	operators
 % $\Omega_m$ as being diagonal, and denote	the	diagonal elements of
 % each	operator as	the	vector ${\bf v}_m$.	The	permutation	symmetric
 % measurements	are	then the measurements such that	under any
 % permutation of the elements of the ${\bf	v}_m$, the set consisting
 % of all the ${\bf	v}_m$ remains unchanged. That is, that under a
 % permutation of their	elements, the ${\bf	v}_m$ merely transform
 % among themselves. Every PSC measurement can be generated	by
 % choosing	a positive operator	$\Omega$ with diagonal vector ${\bf
 % v}$,	and	taking the $\Omega_m$ as all the permutations of $\Omega$
 % (where duplicated operators may be discarded), suitably scaled so
 % that	$\sum_m\Omega_m^2=I$. Thus,	there is a PSC for each	positive
 % operator	$\Omega$.

\vspace{1mm} {\em Theorem:} The classical capacity of a
commutative measurement is the same as that of the equivalent classical 
measurement.

\vspace{1mm} {\em Proof:} Let ${\mathcal M}$ be a commutative 
measurement, and let us denote the basis in which the measurement 
operators of ${\mathcal M}$ are diagonal as $\{|i\rangle\}$.  Note 
that when we use the states $\{|i\rangle\}$ to encode information, the 
behavior of the commutative measurement is exactly the same as the 
equivalent classical measurement.  Consider now the conditional 
probability, $Q(j||\psi\rangle)$, for outcome $j$ given that the 
initial state is the arbitrary state $|\psi\rangle = 
\sum_{i}\alpha_{i}|i\rangle$.  This is
\begin{eqnarray}
  Q(j||\psi\rangle) & = & \mbox{Tr}[E_{j}|\psi\rangle\langle\psi|] 
   = \sum_{i} |\alpha_{i}|^{2} Q(j||i\rangle) .
   \label{revMut2}
\end{eqnarray}
But this is precisely the same expression we would have obtained if we 
had used the mixture $\rho = \sum_{i} |\alpha_{i}|^{2} 
|i\rangle\langle i|$ as the encoding state, instead of the pure state 
$|\psi\rangle$.  Since the expression for the classical capacity can 
be written entirely in terms of the conditional probabilities 
$Q(j||\psi_{i}\rangle)$ (along with the probabilities $P(i)$), this 
means that encoding using any state which is not one of the 
eigenstates $|i\rangle$ renders the same information as encoding using 
a mixture of the eigenstates.  
% It is also therefore equivalent to performing the classical
% measurement and using a probability distribution over the available 
% classical states to encode information, rather than a single state.  
Since using a mixture is never better than using a 
single state, the capacity is achieved by encoding using the 
eigenstates.  In this case the commutative measurement reduces to the 
equivalent classical measurement, and their respective capacities are 
the same.  $\square$

Thus the classical capacity of CPS measurements is given by 
Eqs.(\ref{capCIPM}) and (\ref{capCPM}).

It was shown in the previous section that an ensemble which achieves 
the classical capacity for all unitarily covariant measurements is the 
uniform distribution over any basis.  Using this we calculate the 
resulting classical capacity for irreducible UC measurements in 
Appendix~\ref{appendixB}, which is
\begin{eqnarray}
   C    &=& \ln N - Q(E) - \sum_{k=2}^{N}\frac{1}{k} ,
   \label{covcap}
\end{eqnarray}
where $Q(E)$ is the subentropy of the operator $E$, as defined by 
Jozsa, Robb and Wootters~\cite{JRW}.  The capacity in nats of a 
general UC measurement,
\begin{equation}
  {\mathcal M} = \sum_{n=1}^{N} p_{n} {\mathcal N}_{n} ,
\end{equation} 
where each of the irreducible UC measurements ${\mathcal N}_{n}$ is 
generated by the operator $E_{n}$, is thus
\begin{eqnarray}
   C    &=& \ln N - \sum_{n=1}^{N} p_{n} Q(E_{n}) - \sum_{k=2}^{N}\frac{1}{k} .
\end{eqnarray}
The capacity in nats of the complete unitarily covariant measurements 
(of which there is only one for each dimension $N$), being a special 
case of the above formula, is
\begin{equation}
 C = \ln N - \sum_{k=2}^{N}\frac{1}{k} .
\end{equation}

% \section{Conclusion}
% Included in this conclusion is the conclusion that we have concluded.
% With that we conclude the conclusion.

\section*{Acknowledgments} 
The author would like to thank Gerard Jungman and Howard Barnum for 
helpful discussions, and Howard Wiseman for helpful comments on the 
manuscript.  The author is also grateful both to Vlatko Vedral for 
hospitality during a visit to Imperial College, and Lucien Hardy for 
hospitality during a visit to the Perimeter Institute where part of 
this work was carried out.

\appendix

\section{Terminology}
\label{terms}

The following names are used in the body of the paper to designate
various classes of measurements:

{\bf\em Bare:} Measurements in which every measurement operator is a 
positive operator.  Alternative terms which have been used for these 
kind of measurements are {\em pure} measurements~\cite{DJJ}, 
measurements {\em without feedback}~\cite{FJ}, and {\em square root} 
measurements~\cite{Barnum}.

{\bf\em Classical:} Measurements which can be performed on systems 
which lie within the domain of classical physics.  These are 
measurements in which the all the measurement operators are positive, 
mutually commuting, and also commute with the density matrix 
describing the state being measured.

{\bf\em Commutative:} Measurements in which the all the measurement 
operators are mutually commuting.

{\bf\em Complete:} Measurements in which all the effects are 
proportional to rank-1 projectors~\cite{JRW}.  An alternative name for 
these measurements is {\em maximal strength}.

{\bf\em Finite-Strength:} Measurements in which all the measurement 
operators are of full rank~\cite{FJ}.

{\bf\em Infinite-Strength:} Measurements in which at least one of the 
measurement operators has at least one zero eigenvector.

{\bf\em Incomplete:} Measurements in which at least one of the 
effects is higher than rank one.

{\bf\em Permutation-Symmetric:} Measurements in which, if 
any two of the basis states are permuted, the measurement remains 
unchanged.  That is, under a permutation of basis states, the 
measurement operators merely transform among themselves.

{\bf\em Unitarily Covariant:} Measurements in which, if a unitary 
transformation is applied to the measurement operators, the 
measurement remains unchanged.  That is, under a unitary 
transformation, all the measurement operators transform among 
themselves.

{\bf\em von Neumann:} Measurements in which the all the measurement 
operators are commuting (orthogonal) rank-1 projectors.

\section{Classical capacity of Unitarily Covariant Measurements}
\label{appendixB}
To calculate the classical capacity it is most convenient to use the 
form for the mutual information given in Eq.(\ref{forIc}).  For 
Unitarily covariant measurements the classical capacity is attained by 
an ensemble consisting of the uniform distribution over any basis, so 
we have
\begin{eqnarray}
  C({\cal M}) &=& \sup_\varepsilon \left[ H[P(i)] - \sum_j Q(j)H[P(i|j)] \right]
                  \nonumber \\
              &=& \ln N - \int H [P(i|U)] d\mu(U) \nonumber \\
              &=& \ln N - \int H [N Q(U||i\rangle) P(i)] d\mu(U) \nonumber \\
              &=& \ln N + N\int Q(U||\psi\rangle) \ln Q(U||\psi\rangle) 
              d\mu(U)
              \nonumber \\ 
              &=& \ln N - N H[Q(U||\psi\rangle)]  ,
              \label{expcapcov}
\end{eqnarray}
where we have used the fact that $Q(U||i\rangle)$ is independent of 
$|i\rangle$, and it should be noted that in the second to last 
line, $|\psi\rangle$ simply represents any pure state, and $d\mu(U)$ is the 
unitarily covariant Haar measure over unitary transformations. To continue, 
\begin{eqnarray}
	Q(U||\psi\rangle) &=& \mbox{Tr}[UEU^{\dagger} |\psi\rangle\langle\psi|]
                       = \langle\psi| UEU^{\dagger} |\psi\rangle 
    \nonumber \\      
                      &=& \sum_{j=1}^{N} E_{j} |\langle\psi|U|E_{j}\rangle|^{2} .
\end{eqnarray}
Thus the entropy of the conditional probability density is
\begin{widetext}
\begin{eqnarray}
 H[Q(U||\psi\rangle)] 
    & = & - \int 
            \sum_{j=1}^{N} E_{j}|\langle\psi|U|E_{j}\rangle|^{2}
            \ln\left(\sum_{k=1}^{N}
	        E_{k}|\langle\psi|U|E_{k}\rangle|^{2}\right) 
	        d\mu(U) 
	\nonumber \\
    & = & - \int 
            \sum_{j=1}^{N} E_{j}|\langle\psi|E_{j}\rangle|^{2}
            \ln\left(\sum_{k=1}^{N} 
	        E_{k}|\langle\psi|E_{k}\rangle|^{2}\right) 
	        d\mu(|\psi\rangle) ,
\end{eqnarray}
where $d\mu(|\psi\rangle)$ is the unitarily covariant measure over pure 
states~\cite{Jones}.  Writing $P_{i} = |\langle\psi| 
E_{j}\rangle|^{2}$, this becomes an integral over the uniform measure 
on the probability simplex~\cite{Sykora,pool}, being the volume defined 
by $\sum_{i}P_{i} \leq 1$.  That is
\begin{eqnarray}
   N H[Q(U||\psi\rangle)]	&=&	- N!	
							 \int_{0}^{1}\int_{0}^{1-P_{1}}
						     \!\!\!\!\!\!\!\!\!\!\!\!
                        	 \cdots
       	                     \int_{0}^{1-\sum_{n=1}^{N-2}P_{n}} 
						     \sum_{j=1}^{N} E_{j}P_{j}
						     \ln\left(\sum_{k=1}^{N}	E_{k}P_{k} \right)
 							 \;	dP_{N-1}\cdots dP_{1} ,
   \label{bigint}
\end{eqnarray}
\end{widetext}
where $P_{N}= 1 - \sum_{n=1}^{(N-1)}P_{n}$. Evaluating this integral 
is non-trivial, but it has been solved by Jones~\cite{Jones}, and two 
alternative methods are given by Jozsa, Robb and Wootters~\cite{JRW}. 
The solution is
\begin{eqnarray}
   N H[Q(U||\psi\rangle)]	&=&	-\sum_{k=1}^{N} \left( 
   (E_{k}\ln E_{k}) \prod_{l\not= k} \frac{E_{k}}{E_{k}-E_{l}}
   \right) \nonumber \\
                        & & + \left( \frac{1}{2} + \frac{1}{3} + 
                        \cdots + \frac{1}{N} \right)  \nonumber \\
                        &=& Q(E) + \sum_{k=2}^{N}\frac{1}{k} ,
   \label{gencov}
\end{eqnarray}
where $Q(E)$ is the subentropy of the operator $E$, as defined by 
Jozsa, Robb and Wootters.  It might appear that this blows up when any 
of the eigenvalues of $E$ are equal --- however this is not the case; 
the value of $H[Q(U||\psi\rangle)]$ in the limit as $E_{n}\rightarrow 
E_{m}$, for any $n$ and $m$, remains finite~\cite{JRW}.  In fact, 
$Q(E) \leq S(E), \forall E$.  Combining Eq.(\ref{gencov}) with 
Eq.(\ref{expcapcov}), gives the capacity (in nats) of all covariant 
measurements generated from a single operator $\Omega=\sqrt{E}$.  It 
is shown in Ref.~\cite{JRW} that the maximum value of the right hand 
side of Eq.(\ref{gencov}) is $\ln N$, which is obtained when $E=I/N$.  
Thus the capacity of the covariant measurement generated by 
$\Omega=I/\sqrt{N}$ is zero as required.

For complete measurements, $E_{1}=1$ and $E_{j}=0$ for $j>1$, and the 
integral in Eq.(\ref{bigint}) reduces to
\begin{eqnarray}
	H[Q(U||\psi\rangle)] &=& -(N-1)\int_{0}^{1} 
	(1-P)^{N-2}P\ln\left(P\right) dP \nonumber \\
	& = & \frac{1}{N} 
	      \left( \frac{1}{2} + \frac{1}{3} + \cdots + \frac{1}{N} \right) .
\end{eqnarray}
Combining this with Eq.(\ref{expcapcov}) gives the classical capacity 
(in nats) for the complete unitarily covariant measurements.

\end{document}